\documentclass[twocolumn]{aastex631}

\usepackage{multirow}

\newcommand{\gizmo}{\textsc{gizmo}}
\def\msol{M_\odot}
\def\Ha{H$\alpha$}

\defcitealias{lucchini20}{L20}
\defcitealias{besla12}{B12}
\defcitealias{pardy18}{P18}

\submitjournal{ApJL}

\shorttitle{The Magellanic Stream at 20~kpc}
\shortauthors{Lucchini et al.}
\graphicspath{{./}}

\begin{document}

\title{The Magellanic Stream at 20~kpc:\\A New Orbital History for the Magellanic Clouds}

\correspondingauthor{Scott Lucchini}
\email{lucchini@wisc.edu}

\author[0000-0001-9982-0241]{Scott Lucchini}
\affiliation{Department of Physics, University of Wisconsin - Madison, Madison, WI, USA}

\author[0000-0003-2676-8344]{Elena D'Onghia}
\affiliation{Department of Physics, University of Wisconsin - Madison, Madison, WI, USA}
\affiliation{Department of Astronomy, University of Wisconsin - Madison, Madison, WI, USA}

\author[0000-0003-0724-4115]{Andrew J. Fox}
\affiliation{AURA for ESA, Space Telescope Science Institute, 3700 San Martin Drive, Baltimore, MD, USA}

\begin{abstract}
We present new simulations of the formation of the Magellanic Stream based on an updated first-passage interaction history for the Magellanic Clouds, including both the Galactic and Magellanic Coronae and a live dark matter halo for the Milky Way. This new interaction history is needed because previously successful orbits need updating to account for the Magellanic Corona and the loosely bound nature of the Magellanic Group.
These orbits involve two tidal interactions over the last 3.5~Gyr and reproduce the Stream's position and appearance on the sky, mass distribution, and velocity profile.
Most importantly, our simulated Stream is only $\sim$20~kpc away from the Sun at its closest point, whereas previous first-infall models predicted a distance of $100-200$~kpc. This dramatic paradigm shift in the Stream's 3D position would have several important implications. First, estimates of the observed neutral and ionized masses would be reduced by a factor of $\sim$5. Second, the stellar component of the Stream is also predicted to be $<$20~kpc away. Third, the enhanced interactions with the MW's hot corona at this small distance would substantially shorten the Stream's lifetime. Finally, the MW’s UV radiation field would be much stronger, potentially explaining the \Ha\ emission observed along most of the Stream. Our prediction of a 20~kpc Stream could be tested by searching for UV absorption lines toward distant MW halo stars projected onto the Stream.
\end{abstract}



\section{Introduction} \label{sec:intro}

The Magellanic System is essential to our understanding of the ongoing formation and evolution of the Local Group.
It consists of the two closest massive dwarf galaxies to the Milky Way (MW), the Large and Small Magellanic Clouds (LMC and SMC), and the Magellanic Stream, a massive network of gaseous filaments trailing behind the Clouds (see \citealt{donghia16} for a review).
Significant theoretical and computational effort has gone into furthering our understanding of the origin and formation of the Magellanic Stream, beginning with simple analytical models \citep[e.g.][]{fujimoto77}, eventually including simple hydrodynamics and self-gravity \cite[e.g.][]{moore94,gardiner96}.
Modern high-resolution simulations include live N-body models for both the LMC and the SMC as well as self-consistent hydrodynamics with radiative cooling and star formation (\citealt{besla12}, hereafter \citetalias{besla12}; \citealt{hammer15}; \citealt{pardy18}, hereafter \citetalias{pardy18}; \citealt{wang19}). Most recently, \citeauthor{lucchini20} (\citeyear{lucchini20}, hereafter \citetalias{lucchini20}) showed that including a Magellanic Corona of warm gas surrounding the LMC and SMC can explain the ionized gas component of the Magellanic Stream \citep{fox14}. 

However, as these models have improved, one piece of the puzzle has remained unconstrained: the exact past orbits of the LMC and SMC. Proper motion (PM) measurements for the Magellanic Clouds have become very precise \citep{kallivayalil13,zivick18}, but the total mass of the MW and the LMC are still imprecisely known \citep{bland-hawthorn16,donghia16}. 
The PM measurements favor a first-infall scenario \citep{besla07}, which is supported by the LMC's wake in the dark matter (DM) distribution of the MW halo \citep{conroy21}. The largest uncertainties in the orbits of the Clouds come from hydrodynamical effects including ram pressure and tidal energy losses, which are difficult to include in analytical integrators.
Given the recent indications for the Magellanic Corona \citepalias[see][]{lucchini20}, as well as the need to include the MW's hot circumgalactic medium (CGM), these hydrodynamical effects will play a significant role in the orbital history of the Clouds. Upon the inclusion of the gaseous halos, we find that the Clouds can survive fewer recent interactions than previously thought \citepalias{besla12} if they are to remain separated at the present day. Additionally, evidence for a Magellanic Group \citep[e.g.][]{donghia08,nichols11} suggests that, being loosely bound, its two largest members should expect only a couple of direct interactions within the past $\sim$5$-$7~Gyr.

In this Letter we explore the large-scale structure and location of the Stream resulting from an alternate first-passage interaction history between the Clouds as motivated by the existence of the Magellanic Corona. Our new simulations are consistent with the observed PMs of the Clouds and have dramatic implications on the 3D location of the Magellanic Stream.
In Section~\ref{sec:methods} we outline the methods and initial conditions used in our simulations. In Section~\ref{sec:results} we discuss our main results, and in Section~\ref{sec:discussion} we dissect the significant outcomes and implications of the model.

\section{Methods} \label{sec:methods}

For this work, we used the \gizmo\ massively parallel, multiphysics code \citep{hopkins15,gadget}. \gizmo\ employs a Lagrangian meshless finite-mass (MFM) hydrodynamics scheme that is ideal for simulations with large bulk velocities and large dynamic ranges in density. The MFM scheme provides the ability to track individual fluid elements while still capturing Kelvin-Helmholtz instabilities and shocks \citep{hopkins15}. We used adaptive gravitational softening lengths for gas particles, and softening lengths of 350 and 100~pc for the DM and stellar components, respectively.
Additionally, the default cooling \citep[see Appendix B of][]{hopkins17} and star formation \citep{springel03} routines in \gizmo\ were included.

We do not include metal-line cooling, time-variable ionizing radiation from the MW disk, or UV background radiation in our model, because although these complex mechanisms would influence the Stream's thermal state and ionization level, we expect they would not affect its location, which is the focus of this Letter.
Following \citetalias{lucchini20}, we assume the cold material (the \ion{H}{1} Stream) originates from the disks of the LMC and SMC (the Magellanic ISM), whereas the warm ionized material originates from the Magellanic Corona.

\begin{table*}[t]
	\footnotesize
    \begin{flushleft}
    \caption{Initial and Final Properties of the Galaxies in the Simulation.}
    \label{tab:ics}
    \begin{tabular*}{\textwidth}{l @{\extracolsep{\fill}} ccc}
        \hline \hline
        & MW & LMC & SMC \\ \hline
        $v_{200}$ (km s$^{-1}$) & 166.1 & 92.72 & 45.22 \\
        DM Concentration$^a$ & 12 & 9 & 15 \\
        DM Mass ($\msol$) & $10^{12}$ & $1.8\times10^{11}$ & $1.9\times10^{10}$ \\
        Stellar Mass ($\msol$) & $4.8\times10^{10}$ & $5\times10^9$ & $2.6\times10^8$ \\
        Stellar Scale Length (kpc) & 2.4 & 0.9 & 0.8 \\
        Disk Gas Mass ($\msol$) & $10^{10}$ & $5\times10^9$ & $1.6\times10^9$ \\
        Gas Scale Length (kpc) & 7.0 & 2.8 & 2.0 \\
        Corona Mass ($\msol$) & $10^{11}$ & $8.3\times10^9$ & $-$ \\
        Corona Temp (K) & $2.4\times10^6$ & $2.4\times10^5$ & $-$ \\
        N Particles & $7.8\times10^6$ & $2.6\times10^6$ & $5.3\times10^5$ \\
        Initial Position (kpc) & (0, 0, 0) & (47.36, 546.38, 150.52) & ($-$19.79, 412.29, 183.75) \\
        Initial Velocity (km s$^{-1}$) & (0, 0, 0) & (1.71, $-$99.02, $-$63.73) & (13.43, $-$77.21, $-$80.33) \\\hline
        Sim Position (kpc) & $-$ & (2.0, $-$40.8, $-$31.0) & (13.3, $-$39.3, $-$45.4) \\
        Observed Position (kpc) & $-$ & ($-$1.0, $-$40.9, $-$27.7)$^e$ & (14.9, $-$38.1, $-$44.2)$^e$ \\
        Sim Velocity (km s$^{-2}$) & $-$ & ($-$101.1, $-$275.8, 229.2) & ($-$89.8, $-$300.1, 168.6) \\
        Observed Velocity (km s$^{-1}$) & $-$ & ($-57\pm13$, $-226\pm15$, $221\pm19$)$^e$ & ($18\pm6$, $-179\pm16$, $174\pm13$)$^f$ \\\hline
        Sim PM$^{b,c}$ (mas yr$^{-1}$) & $-$ & ($-2.18\pm0.02\pm0.23$, $0.20\pm0.01\pm0.28$) & ($-1.10\pm0.02\pm0.45$, $-0.93\pm0.02\pm0.49$) \\
        Observed PM$^b$ (mas yr$^{-1}$) & $-$ & ($-1.91\pm0.02$, $0.23\pm0.05$)$^e$ & ($-0.83\pm0.12$, $-1.21\pm0.04$)$^f$ \\
        Sim RV$^d$ (km s$^{-1}$) & $-$ & $262.2\pm6.2\pm12.5$ & $165.8\pm5.0\pm32.0$ \\
        Observed RV (km s$^{-1}$) & $-$ & $262.2\pm3.4^g$ & $145.6\pm0.6^h$ \\
        \hline
    \end{tabular*}
    \end{flushleft}
    \footnotesize{\raggedright \textbf{Notes.} The resultant galaxies have rotation curve peaks of $\sim$240, $\sim$120, and $\sim$65 km s$^{-1}$ at 12, 8.5, and 5 kpc for the MW, LMC, and SMC respectively.}\\
    \footnotesize{$^a$ The DM concentration parameter, $c$, is defined as the ratio of the virial radius, $R_\mathrm{vir}$, to the Navarro-Frenk-White (NFW) scale radius, $R_s$: $R_\mathrm{vir}=cR_s$. The NFW scale radius is then converted into a Hernquist scale radius with $a=R_s\sqrt{2\left(\log(1+c)-c/(1+c)\right)}$.\\
    $^b$ Proper motions given as ($\mu_W$, $\mu_N$) following the convention in \citet{kallivayalil13}.\\
    $^c$ Simulation PM errors are given as ($\mu_{W,N}\pm$err$_\odot\pm$err$_\mathrm{res}$), where err$_\odot$ is the error due to the propagation of the uncertainties in the observed solar velocity and location via the bootstrapping method (solar values and errors from \citealt{kallivayalil13}, section 5). err$_\mathrm{res}$ is an approximation of the variability in kinematics of the Clouds due to small-scale power effects, which change with the numerical resolution. It is computed by measuring the standard deviation of the resultant PMs and RVs in simulations of three different resolutions.\\
    $^d$ Simulation radial velocity errors are given as ($v_\mathrm{rad}\pm$err$_\odot\pm$err$_\mathrm{res}$) with err$_\odot$ and err$_\mathrm{res}$ as defined above in $c$. \\
    $^e$ \citet{kallivayalil13}; $^f$ \citet{zivick18}; $^g$ \citet{vandermarel02}; $^h$ \citet{harris06}}
\end{table*}

\subsection{Initial Conditions} \label{sec:ics}

The simulation presented here contains the same components as those used in \citetalias{lucchini20}, except we have added a live DM halo for the MW.
In brief, those components are galaxies with stellar and gaseous exponential disks embedded in live Hernquist profile DM halos 
following the methods outlined in \citet{springel05}, and gaseous coronae around the LMC and MW. Table~\ref{tab:ics} outlines the parameters used to generate the initial setup of the simulation.

Based on recent measurements \citep{penarrubia16,erkal19}, and the mounting evidence for a Magellanic Group \citep[e.g.][]{donghia08,nichols11}, we include a high-mass LMC in our model. It has a total DM mass of $1.8\times10^{11}~\msol$. Our SMC is consistent with previous works \citepalias{pardy18,lucchini20} with a DM mass of $1.9\times10^{10}~\msol$.
We use the MW model of \citet{donghia20} with a gaseous component added to the disk. The total number of particles used for each galaxy is given in
row 10 of Table~\ref{tab:ics}, and leads to masses per particle ranging from $1.8\times10^5$ to $7.0\times10^5~\msol$ for DM, $4.2\times10^3$ to $1.6\times10^4~\msol$ for stars, and $4.3\times10^3$ to $1.8\times10^4~\msol$ for gas.

Following \citetalias{lucchini20}, we included a Magellanic Corona around the LMC and SMC by extracting the radial density profile from an LMC analog in the Auriga cosmological simulations \citep{grand17}, although we used slightly different selection criteria. Specifically, we selected all gas with $T>2.5\times10^{4}$~K, the lowest temperature at which the galaxy's disk gas was excluded from the selection. The corona was then added around our LMC with its total mass scaled by the ratio between the Auriga LMC analog's total mass and our LMC's total mass, leading to a mass of $8.3\times10^{9}~\msol$ in the Magellanic Corona and a density of $10^{-4}$ cm$^{-3}$ at 50 kpc from the LMC. Finally the corona's temperature was set to $2.4\times10^5$~K, the expected virial temperature of the LMC. These values are also consistent with other recent cosmological simulations \citep{hafen19,jahn21}. Run in isolation for 4~Gyr, the LMC and Magellanic Corona remain stable.

Around the MW, we initially included a gaseous corona following the ``fiducial'' density profile in \citet{salem15}: a $\beta$-profile with $n_0=0.46$ cm$^{-3}$, $r_c=0.35$ kpc, and $\beta=0.559$. The maximum density was capped within 13~kpc from the Galactic Center where the corona overlapped with the gaseous disk. Additionally, the profile exponentially declines for $r>r_\mathrm{vir}=166$~kpc. This initial simulation was unable to reproduce the velocity profile of the Stream (see Section~\ref{sec:results}), so in the final simulation presented here, the total mass was increased by a factor of two to $4\times10^{10}~\msol$, which solved the kinematic discrepancy. This increased the MW halo density at 50~kpc from $1.1\times10^{-4}$ to $2.0\times10^{-4}$~cm$^{-3}$, still consistent with current data (e.g. \citealt{anderson10}; \citealt{li17}).
We acknowledge that the Galactic and Magellanic Coronae included in these simulations are simplified when compared with the complex multiphase intricacies known to exist in circumgalactic media \citep{vandevoort19}.

\subsection{Orbits} \label{sec:orbits}

The orbits used in previous simulations of Magellanic Stream formation were established without the Galactic or Magellanic Coronae included \citepalias{besla12,pardy18}.
While the Cloud orbits will clearly require modification due to the changes in total masses of the galaxies due to these added components, the increased friction and ram pressure that the Clouds experience as they move through these media also play a significant role.
Therefore, to match the observed positions and velocities of the Magellanic Clouds while including the Galactic and Magellanic Coronae in our simulations, we need to determine an alternate orbital history with fewer recent interactions between the Clouds.

To do this, we analytically integrated the orbits of the LMC, SMC, and MW backward in time starting from their present-day observed positions and velocities. We included radially extended Hernquist DM halos including dynamical friction and represented the stellar and gaseous disks as point particles. By varying the present-day velocities within the observed errors, we obtained a suite of 1458 possible orbits. We filtered these to select first-passage orbits with multiple interactions between the Clouds. These orbits are generally consistent with the results found with previous analytic integrations after accounting for differences in LMC mass and live versus static MW DM halos \citep{besla07,kallivayalil13,garavito-camargo19}.
We then chose 10 of these orbits sampling the parameter space with varying morphologies to run in full \textit{N}-body hydrodynamical simulations.
Due to the effects of tidal stripping, ram pressure, and friction, the initial conditions required modification to match the present-day observations of the Clouds. For 9 of the 10 chosen orbits, we were unable to modify them such that they reproduced the present-day observations.
However, after a few iterations on one of the orbits, we found a solution that could match the Clouds' observed kinematic properties, which we present in this Letter. The initial positions and velocities used for the orbits in this simulation are given in
rows 11 and 12 
of Table~\ref{tab:ics}.

Given the backward-integrated orbits, we had to choose when to begin the \textit{N}-body simulation. We chose the apocenter between the Clouds after their second interaction (3.5~Gyr ago). While the Clouds would have another encounter if we integrate these orbits further back in time, this interaction would occur 7 Gyr ago (5.5 Gyr before their next interaction) at a distance of 0.9 Mpc away from the MW. Any gas stripped in this interaction would be tidally thrown out to great distances from the Clouds and become too diffuse and distant to contribute to the Stream today (see material stripped from the first interaction in Figure~2 of \citetalias{pardy18}).

\subsubsection{Comparison with Previous Orbits} \label{sec:comparison}

In the previous orbital model of \citetalias{besla12} and \citetalias{pardy18} (\citetalias{besla12}'s ``Model 2''), the LMC and the SMC experience three interactions in isolation (without any MW influence) over $\sim$6~Gyr. They are then rotated into the correct orientation, placed at the virial radius of the MW, and allowed to continue evolving until they reach their present-day positions ($\sim$1~Gyr). During this infall they experience an additional direct collision that forms the Magellanic Bridge.

\begin{figure*}[t]
\includegraphics[width=6.65in]{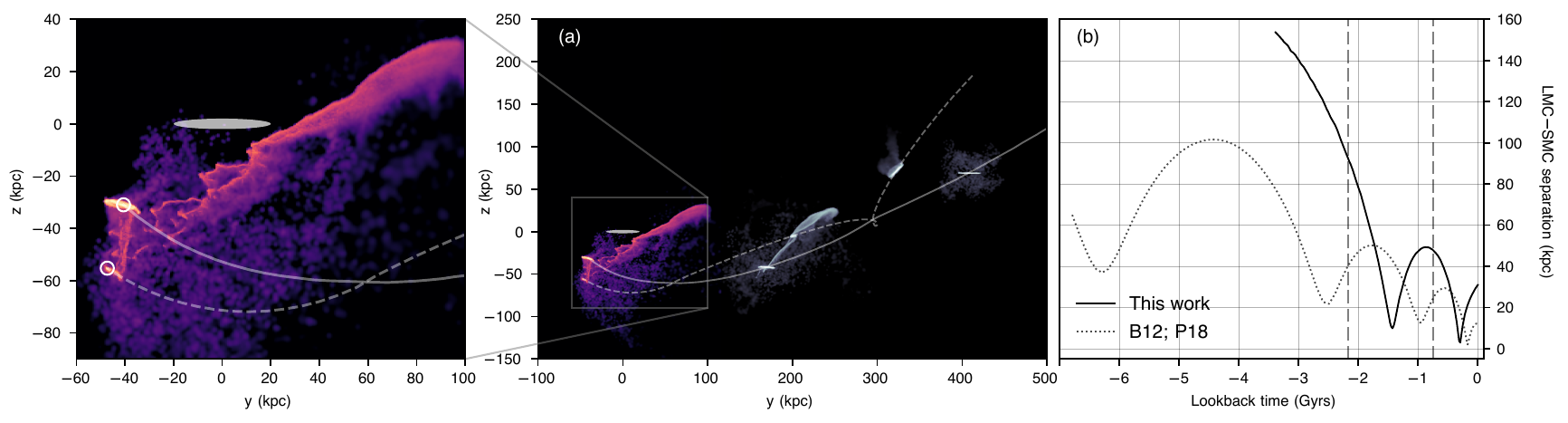}
\caption{The orbital history of the Magellanic Clouds. Panel (a) shows the Magellanic Stream in Cartesian coordinates at three different times during the Clouds' infall. The present-day Stream is shown in color. The MW disk is at the origin and denoted by the gray shaded oval (the sun is located at $(x,y,z)=(-8.3,0,0.027)$~kpc). The solid and dashed lines represent the past orbital trajectories of the LMC and SMC, respectively. A zoomed inset of the present-day Stream can be seen in the panel on the left with the locations of the Clouds in the simulation marked with circles. The color of the gas represents plane-projected density on an arbitrary scale with higher densities represented as lighter colors. Panel (b) shows the distance between the center of masses of the LMC and SMC in kpc as they orbit around each other and fall in toward the MW. The model presented here is compared with previous works (shown as a dotted line; \citetalias{besla12,pardy18}). Present day is on the right side of the plot ($t=0$), and the Clouds' initial state is on the left side ($t=-3.5$~Gyr for the present work).
Vertical dashed lines denote the times of the two past images in panel (a) (shown in gray scale). An animation of this figure is available online. It is 9 s long and shows the evolution of the Clouds and the Stream over the past 3.5~Gyr.}
\label{fig:cartesian}
\end{figure*}

In the model presented here, all the interactions between the Clouds and the MW occur in a single, self-consistent simulation lasting 3.5~Gyr. The LMC and the SMC have two interactions, the second of which has a very low impact parameter and forms the Magellanic Bridge. The main differences between these two orbital models are (see Figure~\ref{fig:cartesian}b):
\begin{itemize}
    \item the number of interactions (4 versus 2),
    \item the length of the simulation (7 versus 3.5~Gyr),
    \item the maximum separation between the Clouds (100 versus 150~kpc), and
    \item the sense of the SMC's orbit around the LMC (see Section~\ref{sec:discussion} and Figure~\ref{fig:schematic}).
\end{itemize}
While there have been many other proposed orbital models of the interactions between the Clouds \citep[e.g.][]{ruzicka10,diaz11,guglielmo14}, none have looked at this explicit combination, especially when considering a first-passage scenario with the inclusion of an MW CGM and the Magellanic Corona.

\section{Results} \label{sec:results}

The simulation ran for 3.46~Gyr, when the positions and velocities of the Clouds matched current observations just after their first pericentric passage. When the LMC and SMC are at their present-day sky positions in our model, they are at distances of $52.1\pm1.7$~kpc and $78.0\pm7.8$~kpc respectively (with errors calculated as $\mathrm{err}_\mathrm{res}$ discussed in footnote c of Table~\ref{tab:ics}; $\mathrm{err}_\odot$ are negligible), with proper motions and radial velocities listed in Table~\ref{tab:ics}.
The kinematics of the Clouds are fully consistent with the observed values: within $1\sigma$ for the proper motions and radial velocity for both the LMC and the SMC (with the exception of $\mu_W$ for the LMC at 1.18$\sigma$). Additionally, the relative velocity between the Clouds (66 km s$^{-1}$) matches within 2$\sigma$.

The present-day disk gas masses\footnote{These masses were calculated in physical 3D space by summing the particle masses within spheres centered on each galaxy with diameters 13.5 kpc for the LMC, and 5.5 kpc for the SMC.} for the LMC and SMC are $4.4\times10^8$ and $3.3\times10^7~\msol$ respectively (with peak column densities of $10^{21.8}$ and $10^{21.4}$ cm$^{-2}$). A full exploration of the structure of the Clouds themselves will be performed in an upcoming paper with more complete star formation and feedback routines (Lucchini, S. et al. 2021, in preparation).

\begin{figure*}[t]
\plotone{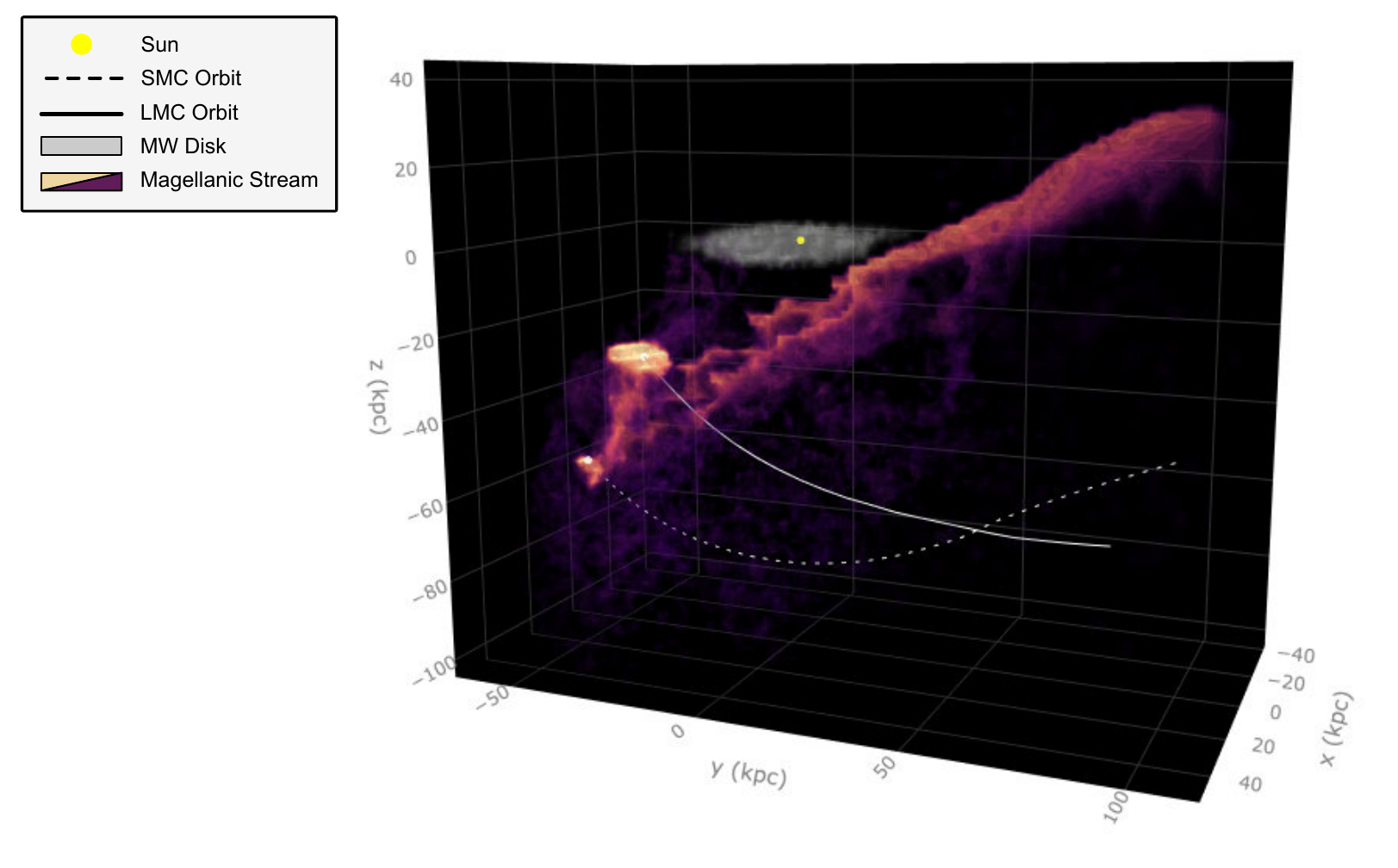}
\caption{A 3D model of the Magellanic Stream. The orbits of the LMC and SMC are shown as solid and dashed lines, respectively. The MW disk is shown in gray with the location of the Sun marked with a yellow sphere. An interactive version of this figure is available online.
\label{fig:3d}}
\end{figure*}

Throughout their 3.5~Gyr interaction history, a trailing Stream is formed through tidal interactions between the Clouds. The orbital paths of the LMC and SMC along with the resultant Stream at the present day are shown projected onto the $y$--$z$ plane relative to the MW disk in Figure~\ref{fig:cartesian}a. The LMC and SMC experience two close encounters shown as minima in Figure~\ref{fig:cartesian}b. Their first interaction, 1.4~Gyr ago with an impact parameter of 9.9~kpc, provides the tidal forces necessary to strip material from the SMC to create the bulk of the \ion{H}{1} Stream. Their second interaction, 295~Myr ago, has a significantly lower impact parameter of 3.0~kpc, and this direct collision forms the Magellanic Bridge.
The present-day position of the Magellanic System can also be seen in 3D in Figure~\ref{fig:3d}. 

To test the dominant stripping mechanism in our model, we ran two additional simulations: one consisting of just the MW, its hot corona, and the SMC (without the LMC), and another with the MW, its hot corona, and the LMC with the Magellanic Corona (without the SMC). When the SMC alone passes through the MW's CGM, we find only 8\% of the total gas mass stripped from the full model. When the LMC alone passes through, we find negligible stripping of gas from the LMC, but the Magellanic Corona still sees 95\% of the stripping of the full model. Therefore, we conclude that the neutral Stream is stripped through tidal forces (consistent with previous findings; e.g. \citealt{salem15}), whereas the ionized component of the Stream (originating in the Magellanic Corona) is stripped mostly through ram pressure against the MW hot corona, even during the first passage.

\begin{figure*}
\centering
\includegraphics[width=5.5in]{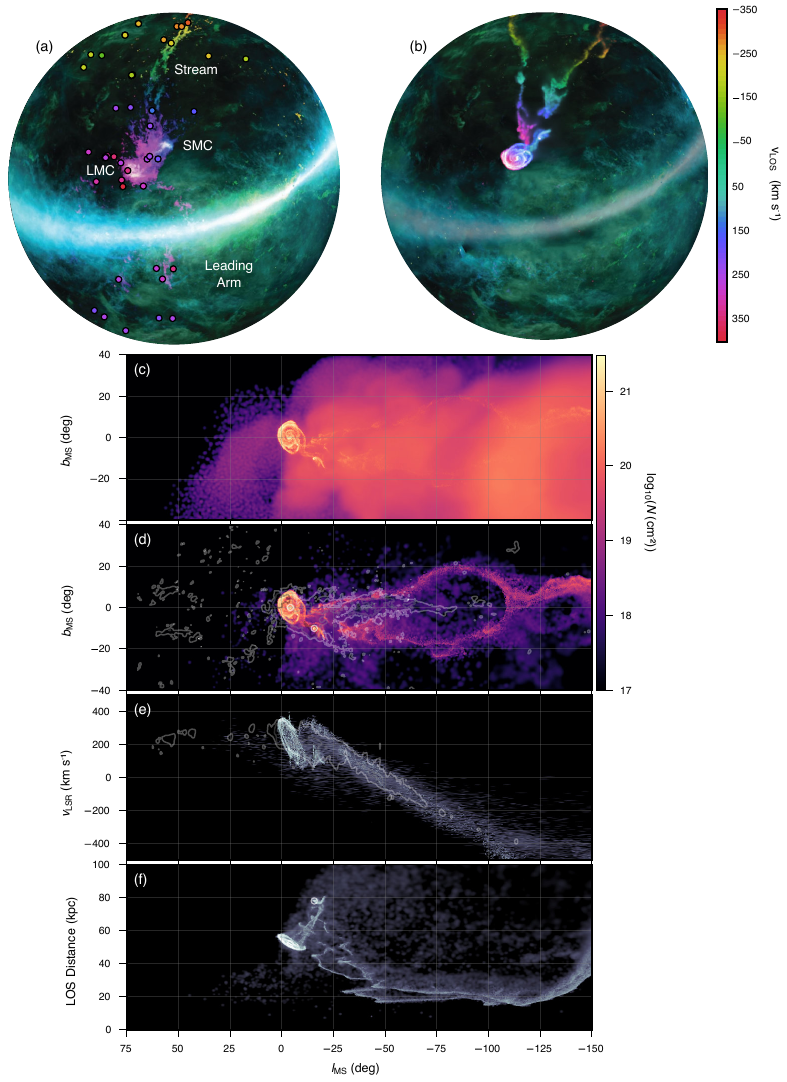}
\caption{Properties of the Magellanic Stream produced in our simulations. Panels (a) and (b) show the observed and simulated Stream respectively in zenithal equal-area coordinates with line-of-sight velocity indicated by the color scale and the relative gas column density indicated by the brightness. The \ion{H}{1} data in Panel (a) are from the GASS survey \citep{mcclure-griffiths09} with the points showing sight  lines with UV-absorption-line observations from the Hubble Space Telescope \citep{fox14} colored by their line-of-sight velocity. Panel (c) shows the column density of the total gas in the Stream (including the ionized Magellanic Corona and neutral Magellanic disk components) in Magellanic Coordinates ($l_\mathrm{MS}$ and $b_\mathrm{MS}$). Panel (d) only shows the neutral gas originating in the disks of the LMC and SMC compared to the observed data from \citet{nidever10} shown in contours (black, gray, and white correspond to $10^{21}$, $10^{20}$, and $10^{19}$~cm$^{-2}$). The centroids of the LMC and SMC stellar disks are marked by white circles. Panel (e) shows the local standard of rest (LSR) velocity gradient along the Stream with data shown as contours \citep{nidever10}. Panel (f) shows the line-of-sight distance to the gas in the Stream along its length with the centroids of the Clouds marked with circles. Note that the bulk of the simulated Stream is significantly closer to us than the Magellanic Clouds are ($\sim$20~kpc vs. $\sim$60~kpc).}
\label{fig:magellanic}
\end{figure*}

As in \citetalias{lucchini20}, the bulk of the mass of the Magellanic Stream is composed of ionized gas originating in the Magellanic Corona.
However, the distribution of the ionized Magellanic Corona gas on the sky is substantially different in our new model (compare Figure~\ref{fig:magellanic}(c) with \citetalias{lucchini20} Figure~2(a)), but the neutral Stream's appearance on the sky in our model is generally consistent with previous models and observations (Figure~\ref{fig:magellanic}(a),  (b), (d)). While the simulated Stream is longer than observed and slightly offset spatially, the morphology of the Stream in this model more accurately reproduces the turbulent, filamentary nature of the data. Due to its interactions with the Magellanic Corona and the MW's hot CGM, instabilities fragment and distort the Stream, leading to a significant improvement in its appearance and morphology when compared to models that do not include these gaseous components \citepalias[e.g.][]{pardy18}.

The velocity profile of our simulated Stream also matches observations (see Figure~\ref{fig:magellanic}e). Previous models found a velocity gradient too shallow when compared to the data \citepalias{lucchini20}, but this is resolved in our new model. As stated in Section~\ref{sec:ics}, the density profile and total mass of the MW hot corona were increased by a factor of 2 over the \citet{salem15} values to better match the velocity gradient along the Stream. If the MW's hot halo is not massive enough, the stripped gas from the Clouds is accelerated toward the MW and in some cases ends up with greater velocities than the LMC and SMC themselves. A higher gas density around the MW provides the ram   pressure forces to slow down the trailing Stream to match the observed velocity gradient. This is in contrast to previous works that placed upper limits on the MW coronal mass such that a Leading Arm can form (\citealt{tepper-garcia19}; \citetalias{lucchini20}).

Additionally, as seen in Figure~\ref{fig:magellanic}d, this model does not self-consistently reproduce the Leading Arm gas. This is because the Clouds only have two close encounters 
so there is not enough time for gas to be tidally thrown ahead of the Clouds in their orbits. This is in contrast to previous studies (\citealt{tepper-garcia19}; \citetalias{lucchini20}) where the lack of a Leading Arm was due to the MW hot coronal density being too high. In this new model, even without an MW hot corona, a Leading Arm is not formed. The true nature of the Leading Arm is one of the biggest outstanding questions in the Magellanic System \citep{donghia16}. While a Magellanic origin is supported by the kinematics \citep{putman98} and the metallicities \citep[although they vary with position;][]{fox18}, several works have proposed alternative, non-Magellanic sources. \citet{hammer15} and \citet{tepper-garcia15} suggested the Leading Arm structures could be remnants from dwarf spheroidal satellites of the MW whose gas has been stripped from the MW hot corona, and a non-Magellanic explanation for the Leading Arm remains a possibility.

\begin{figure*}
    \centering
    \includegraphics{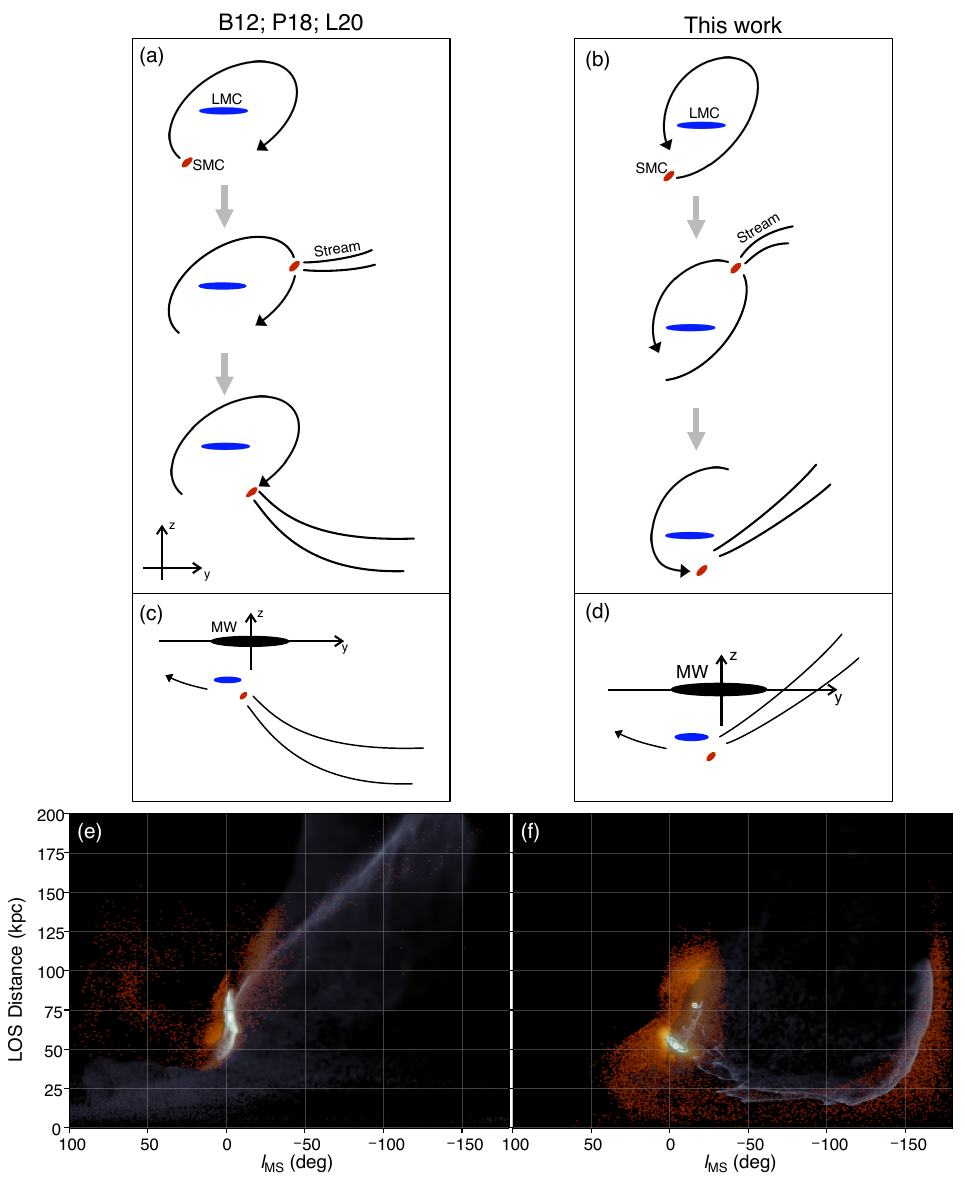}
    \caption{Orbital schematics and distance to the Magellanic Stream compared for two different models. The left column (panels (a), (c), and (e)) shows the clockwise orbit of previous works \citepalias{besla12,pardy18,lucchini20}, while the right column (panels (b), (d), and (f)) shows the counterclockwise orbit used in the model presented here. In panels (a) and (b), the orbital path of the SMC (red) around the LMC (blue) is shown at three different times in the $y$--$z$ plane: before the interaction (top), at apoapsis when the Stream material is stripped out of the SMC (middle), and at their present-day orientation (bottom). Panels (c) and (d) show the present-day positions of the Clouds and the Stream in the two models with respect to the MW, again in the $y$--$z$ plane. The arrow shows the direction of motion of the Clouds around the MW. Panels (e) and (f) both show line-of-sight distance to the gas in the simulated Stream (gray) and the stars in the Stream (orange). In previous orbital models (left column), the Stream stretches out and away from the MW leading to distances of 100--200 kpc. Whereas in the new orbital history presented here (right column), the Stream is stripped during the SMC's counterclockwise motion around the LMC and then pushed into place through the ram pressure and frictional forces of the MW's CGM leading to distances of as little as $\sim$20 kpc.}
    \label{fig:schematic}
\end{figure*}

The most notable implication of this new model is that the Stream is significantly closer to us than previously thought (Figure~\ref{fig:magellanic}f). While some past models have predicted a close Stream (via multiple passages around the MW; e.g. \citealt{moore94,diaz12}), all previous first-passage orbits have resulted in the Stream flowing behind the Clouds out to distances of $100-200$~kpc or greater (see Figure~\ref{fig:schematic}e; \citetalias{besla12,pardy18,lucchini20}).
The first-passage model presented here forms a tidal Stream that reaches as close as $\sim$20~kpc away from the Sun with a column density-weighted average distance of 24.7~kpc (between Magellanic longitudes of $-25^\circ$ and $-150^\circ$).
See Section~\ref{sec:discussion} for an in-depth discussion of this finding.

We emphasize that in all our simulations that formed a Stream (7 of the 10 selected from the backward integration, see Section~\ref{sec:orbits}), that Stream was $<$50~kpc away from us. While the distances and kinematics of the Clouds, and the Stream morphology, vary greatly between these individual simulation runs, the finding that the Stream remains nearby is a robust prediction of our first-infall models.
Additionally, we have run several convergence tests at various numerical resolutions, and despite minor differences on small scales, all runs produced accurate positions and velocities for the Clouds, and all predicted a nearby Stream with total ionized and neutral masses consistent with observations. 

\section{Discussion and Conclusions} \label{sec:discussion}

The new first-passage interaction history of the Magellanic Clouds presented here leads to a dramatically different 3D spatial positioning of the Stream than previous models. Previous first-passage, tidal simulations led to a Stream extending away from the Clouds out to distances upward of 200~kpc, whereas our new model results in the Stream angling up toward the MW reaching as close as 20~kpc to the Sun.
While there are many differences between previous models and the model presented here (see Section~\ref{sec:comparison}), two differences in particular lead to this dramatic shift in positioning of the Stream:
\begin{enumerate}
    \item a qualitative difference in the SMC's orbit around the LMC, and
    \item the inclusion of the Galactic and Magellanic Coronae.
\end{enumerate}

First, the orientation of the SMC's orbit around the LMC is qualitatively different than the most recent previous models \citepalias{besla12,pardy18,lucchini20}. To compare these different orbital histories, we need to analyze the orbits in a consistent coordinate system, so we will discuss the relative motion of the SMC around the LMC when viewed projected onto the $y$--$z$ plane as defined in Figure~\ref{fig:cartesian} (relative to the MW disk, which is in the $x$--$y$ plane with the Sun located at $(x,y,z)=(-8.3,0,0.027)$~kpc)\footnote{Note that this is the same perspective as Figure~3 in \citetalias{besla12} but is not consistent with the coordinate systems in \citetalias{besla12}'s Figure~2, \citetalias{lucchini20}'s Extended Data Figure~2, or \citetalias{pardy18}'s Figure~2. Because of rotations performed before the Clouds fall into the MW potential in these models, the $y$--$z$ perspective discussed above is approximately equivalent to mirroring these figures across the $y$-axis.}. When viewed from this perspective, our model has the SMC on a \textit{counterclockwise} orbit around the LMC, whereas in the \citetalias{besla12} model, the SMC rotates around the LMC \textit{clockwise}.
In the \textit{clockwise} orbit, the Stream is tidally thrown out in the $+y$ direction with a velocity in the $-z$ direction, leading to it stretching away from the MW disk. Whereas in the \textit{counterclockwise} orbit, the Stream is still tidally stripped in the $+y$ direction, but its velocity is in the $+z$ direction. This leads to a Stream angled up in the $+y$ and $+z$ directions, resulting in low line-of-sight distances to the Sun. A schematic of these two orbital orientations is shown in Figure~\ref{fig:schematic}(a)-(d).

Second, the inclusion of the Galactic and Magellanic Coronae is crucial. While previous parameter space searches of orbits for the Clouds would possibly have explored this orbital configuration in the past \citep[e.g.][]{ruzicka10,guglielmo14}, none included the Galactic and Magellanic Coronae. These are key elements as the ram pressure and friction from the MW's CGM are able to ``push'' the Stream into its present-day position and velocity. Without the MW's gaseous halo, the Stream would collide with the MW disk before the Clouds reach their present-day positions. Moreover, the Magellanic Corona is required to shield the neutral Stream from the intense forces and pressures as it moves through high-density regions of the Galactic corona \citetalias{lucchini20}.

A Stream reaching $\sim$20\,kpc has a number of implications.
First, the total observed mass (neutral and ionized) of the Stream would be reduced. Total mass estimates of the Stream have previously assumed the Stream has a similar distance as the Clouds (55~kpc), but they include a scale factor of $(d/55~\mathrm{kpc})^2$ which equals 0.2 for $d=24.7$~kpc (the column density-weighted average distance in the simulation). This leads to new values of $9.7\times10^7~\msol$ of neutral gas \citep{bruns05} and $4\times10^8~\msol$ of ionized gas \citep{fox14}, although this ionized gas mass should be considered as a lower limit as the spatial extent of the Stream's ionized phase may be significantly greater than the area on the sky that has currently been explored, and the ionized gas may be multiphase \citep{fox14}. Summing the masses of all gravitationally unbound particles that fall in the region of the trailing Stream in our simulation gives values of $2.0\times10^8~\msol$ (neutral) and $3.2\times10^9~\msol$ (ionized) for our model of the Stream, in good agreement with the observations.
Second, as shown in Figure~\ref{fig:schematic}f, the stellar component of the Stream is also nearby, with stars predicted at $d\lesssim20$~kpc (with a mean surface brightness of 31~mag arcsec$^{-2}$). Previous works have predicted that any stars associated with the Magellanic System are at large distances \citepalias{besla12,pardy18}, but our new predictions suggest that continued searches for the Stream's stellar component are worthwhile \citep{zaritsky20}.
Third, the interaction between the Stream and the MW CGM will be enhanced due to its proximity, because of the higher MW corona density. At a closer distance, the Stream would be closer to pressure equilibrium with the MW CGM, which could help explain its multiphase nature \citep{wolfire95} and the high number of ``head-tail'' clouds seen in high-resolution \ion{H}{1} observations \citep{for14}. It could also lead to dramatically shorter lifetimes for the Stream in the future \citep{murali00,bland-hawthorn07}. Indeed, due to its angle of approach with respect to the MW, the Stream may even directly collide with the MW disk within the next $\sim$50~Myr. This moves the timescale of gas accretion onto the disk up by a factor of $\sim$10 from previous predictions. The new distance to the Stream therefore implies a factor of $\sim$8 less gas accreting onto the MW $\sim$10 times earlier than previously thought, resulting in approximately the same infall rate of $\sim$4--7\,$\msol$~yr$^{-1}$ as derived before \citep{fox14}.
Fourth, a closer Stream should be subject to a more intense UV radiation field from the MW and hence significantly brighter in \Ha. This could explain the high observed \Ha\ emission from the Stream \citep{bland-hawthorn13, bland-hawthorn19, barger17}; at a distance of 20~kpc, the Galactic UV background could lead to \Ha\ emission as high as $150-300$~mR \citep{tepper-garcia15}, in agreement with observed levels along most of the Stream. However, this increased radiation still cannot explain the extremely high \Ha\ emission seen in the region of the Stream under the South Galactic Pole, which may indicate a recent Seyfert flare from the Galactic Center \citep{bland-hawthorn19}. The enhanced radiation field intensity at $d$=20 kpc would also affect the ionization level inferred from UV metal-line studies of the Stream \citep{fox14,fox20}; however, the ionized/neutral ratio of $\sim$3:1 is distance independent since both the neutral and ionized masses scale as $d^2$.

There are currently no observational distance constraints on the Stream, although parts of the Leading Arm have constraints of $<20-30$~kpc \citep{mcclure-griffiths08,price-whelan18,antwi-danso20}. To test our prediction of a 20~kpc Stream, UV or optical spectroscopic studies could be performed to look for absorption at Magellanic velocities toward distant MW halo stars, such as blue horizontal branch stars with distances from Gaia. \citet{lehner11} found no UV absorption at Magellanic velocities when looking at 28 halo stars out to $z$-distances of 12.6~kpc, but searches to larger distances using new stellar catalogs are needed. This offers a pathway for confirming our prediction of a nearby Stream.

\begin{acknowledgments}
We thank the anonymous referee for their insightful comments. We thank Bart Wakker for useful discussions and Jay Gallagher and Sne\u{z}ana Stanimirovi\'c for comments on the manuscript. This work was partially supported by HST grant HST-AR-16363.001-A. Additional support was provided by NASA under Award No. RFP21\_4.0 issued through Wisconsin Space Grant Consortium, and the the University of Wisconsin - Madison Office of the Vice Chancellor for Research and Graduate Education with funding from the Wisconsin Alumni Research Foundation. This research was performed using the compute resources and assistance of the UW-Madison Center For High Throughput Computing (CHTC) in the Department of Computer Sciences. The CHTC is supported by UW-Madison, the Advanced Computing Initiative, the Wisconsin Alumni Research Foundation, the Wisconsin Institutes for Discovery, and the National Science Foundation, and is an active member of the Open Science Grid, which is supported by the National Science Foundation and the U.S. Department of Energy's Office of Science.
\end{acknowledgments}

%

\vspace{5mm}


\software{Astropy (\url{https://www.astropy.org/}; \citealt{astropy:2018}),
            \gizmo\ (\url{http://www.tapir.caltech.edu/~phopkins/Site/GIZMO.html}; \citealt{hopkins15,gadget}),
            Matplotlib (\url{https://matplotlib.org/}), 
            Plotly (\url{https://plotly.com/}),
            pygad (\url{https://bitbucket.org/broett/pygad/src/master/}; \citealt{pygad})
          }



\newpage
\bibliography{references}{}
\bibliographystyle{aasjournal}

\listofchanges

\end{document}